\documentstyle[prd,floats,psfig,aps]{revtex}
\begin{document}
\title{Perturbations of the Kerr spacetime in horizon penetrating coordinates}
\author{
Manuela Campanelli$^{1}$,
Gaurav Khanna$^{2,3}$,
Pablo Laguna$^{2,4}$,
Jorge Pullin$^{2}$,
Michael P. Ryan$^{5}$}

\address{1. Albert-Einstein-Institut,
Max-Planck-Institut f{\"u}r Gravitationsphysik,\\
Am M\"uhlenberg 1, D-14476 Golm, Germany}
\address{2. Center for Gravitational Physics and Geometry, 
Department of Physics,\\ The
Pennsylvania State University, 104 Davey Lab, University Park, PA
16802.}
\address{3. Natural Science Division,\\
Southampton College of Long Island University,\\
Southampton NY 11968.}
\address{4. Department of Astronomy and Astrophysics, 
The Pennsylvania State University,\\
525 Davey Lab, University Park, PA 16802.}
\address{5. Instituto de Ciencias Nucleares, UNAM\\
Apartado Postal 70-543 M\'exico 04510 D. F., M\'exico.}
\maketitle
\begin{abstract}
We derive the Teukolsky equation for perturbations of a Kerr spacetime
when the spacetime metric is written in either ingoing or outgoing 
Kerr--Schild form. We also write explicit formulae for setting up the 
initial data for the Teukolsky equation in the time domain in terms of a
three metric and an extrinsic curvature. The motivation of this work
is to have in place a formalism to study the evolution in the 
``close limit'' of two recently proposed solutions to the initial 
value problem in general relativity that are based on Kerr--Schild 
slicings. A perturbative formalism in horizon penetrating coordinates 
is also very desirable in connection with  numerical relativity
simulations using black hole ``excision''. 
\end{abstract}
%\vspace{-10.7cm} 
%\begin{flushright}
%\baselineskip=15pt
%CGPG-00/10-01  \\
%AEI-2000-062\\
%gr-qc/0010034\\
%\end{flushright}
%\vspace{9.7cm}

\section{Introduction:}

There is considerable current interest in studying the collision of
two black holes, since these events could be primary sources of
gravitational waves for interferometric gravitational wave detectors
currently under construction. On the theoretical side, it is expected
that the problem of colliding two black holes will be tackled by some
combination of full numerical and semi-analytical methods. The first
three dimensional collisions of black holes are starting to be
numerically simulated, albeit with very limited resolution and grid
size. Long time stability of the codes is also an issue. It is
therefore of interest to have at hand approximate results which in
certain regimes could be used to test the codes. Among such
approximation methods is the ``close limit'' approximation \cite{PuJa}
in which the spacetime of a black hole collision is represented as a single
distorted black hole. This approximation has been used  successfully
to test codes for the evolution of black hole space-times that are
axisymmetric and tests are under way for ``grazing'' inspiralling
collisions \cite{inspiralpert}. Another realm of application of 
perturbative calculations is to provide ``outer boundaries'' and to extend
the reach of Cauchy codes into the radiation zone far away from the 
black holes, as was demonstrated in \cite{grandlychallenged2}.
Finally, perturbative codes can be used after
a full non-linear binary black hole code has coalesced the holes
to continue the evolution in a simple and efficient fashion
as was demonstrated in \cite{abrahams,lazarusMisner}.

The perturbative approach requires specifying both a background metric
and a coordinate system when performing calculations. For evolutions in the
time domain such as the ones we are considering, one also has to specify an
initial slice of the spacetime. In all perturbative evolutions
performed up to now the background spacetime has either been the
Schwarzschild solution in ordinary coordinates or the Kerr spacetime
in Boyer--Lindquist coordinates. These backgrounds are adequate for
instance, for the study of the evolution of ``close limits'' of the
Bowen--York \cite{BoYo} and ``puncture'' \cite{BrBr} families of
initial data, which reduce in the ``close limit'' to those background
space-times.

There have been two recent proposals for alternative families of
initial data that have some appealing features
\cite{HuMaSh,winicour}. Both these proposals are based on the use of
the Kerr--Schild form of the Schwarzschild (or Kerr) solutions to
represent each of the black holes in the collision. Some of these
solutions do not have an obvious close limit in which they yield
a single black hole, although the close limit can be arranged with
a simple modification of the original proposal.
In the case in which the close limit exists, the initial data 
appear  as a perturbation of the Schwarzschild
or Kerr spacetime, but in Kerr--Schild 
coordinates. If one wishes to evolve perturbatively the spacetime,
this requires having the perturbative formalism set up on a
background spacetime in Kerr-Schild coordinates. To our knowledge,
this has never been done in the past.

The use of Kerr--Schild coordinates appears quite desirable in 
the context of numerical evolutions of black holes. The coordinates
penetrate the horizon of the holes without steep gradients in the metric
components. This makes them amenable to the numerical technique of 
singularity excision, which is sometimes viewed as the key to long term
binary black hole evolutions such as the ones needed for gravitational
wave data analysis purposes \cite{agave}. 

Since the Kerr--Schild coordinates are horizon-penetrating, developing
a perturbative formalism in these coordinates could, in principle, allow 
the  study of perturbations arbitrarily close to the horizon and even
inside the horizon. The traditional perturbative formalisms, based on
Schwarzschild and Boyer--Lindquist (in the Kerr case) coordinates
cannot achieve this. Having a perturbative formalism that works close
to the horizon is desirable in the context of the recently introduced
``isolated horizons'' \cite{isho}. One could exhibit perturbatively
the validity of several new results that are emerging in such context.

In this paper we will develop perturbative equations in Kerr--Schild
coordinates, taking advantage of the fact that the Teukolsky formalism
is coordinate invariant. We will end by constructing a perturbative
equation that is well behaved inside, across and outside the horizon.

The organization of this paper is as follows. In section II we review
the Teukolsky formalism (including the setup of initial data for
Cauchy evolution). In section III we derive the Teukolsky
equation in horizon penetrating Eddington--Finkelstein
coordinates and display a numerical evolution of it. 
In an appendix we discuss the equation in outgoing 
Eddington--Finkelstein coordinates.

\section{The Teukolsky equation}

Fortunately, the perturbative formalism due to Teukolsky \cite{Te} is
amenable to a reasonably straightforward (but computationally
non-trivial) change of background. The Teukolsky formalism is based on
the observation that the Einstein equations written in the
Newman--Penrose formalism naturally decouple in such a way that one
obtains an equation for the perturbative portion of the Weyl
spinor. In the notation and conventions of Teukolsky (which in turn
follows those of Newman and Penrose), the resulting equation (in
vacuum) is,
\begin{eqnarray}
\left[\left(\Delta+3\gamma-\gamma^*+4 \mu+\mu^*\right)
\left(D+4\epsilon-\rho\right)-\left(\delta^*-\tau^*+\beta^*+3\alpha+4\pi
\right)\left(\delta-\tau+4\beta\right)-3\psi_2\right]\psi_4=0, \label{teukor}
\end{eqnarray}
where the quantities in brackets  are computed using the background
geometry and $\psi_4$ is a first order quantity in perturbation
theory. A similar equation can be derived for the $\psi_0$ component
of the Weyl tensor. We will not concentrate on this equation, however,
since it has not proven as useful for evolutions in the time domain
\cite{AnKrLaPa}. 

If we now particularize to a background spacetime given by the Kerr metric in
Boyer--Lindquist coordinates,
\begin{equation}
ds^2=\left(1-2Mr/\Sigma\right) dt^2+\left(4Mar\sin^2
\theta/\Sigma\right)dtd\varphi -\left(\Sigma\over \triangle\right) dr^2 -
\Sigma d\theta^2
-\sin^2\theta\left(r^2+a^2+2Ma^2r\sin^2\theta/\Sigma\right) d\varphi^2,
\end{equation}
where $\Sigma=r^2+a^2\cos^2\theta$ and $\triangle=r^2-2Mr+a^2$ (and
should not be confused with the Newman--Penrose quantity
$\Delta=n^\mu\partial_\mu$)
and considers the Kinnersley tetrad,
\begin{eqnarray}
l^\mu&=&\left[\left(r^2+a^2\right)/\triangle,1,0,a/\triangle\right],\\
n^\mu&=&\left[r^2+a^2,-\triangle,0,a\right]/(2\Sigma),\\
m^\mu&=&\left[ia\sin\theta,0,1,i/\sin\theta\right]/(\sqrt{2}(r+ia\cos\theta)),
\end{eqnarray}
for which the background Newman--Penrose quantities are,
\begin{eqnarray}
\rho&=&-1/(r-ia\cos\theta), \\
\beta&=&-\rho^* \cot\theta/(2\sqrt{2}),\\
\pi&=&ia\rho^2 \sin\theta/\sqrt{2},\\
\tau&=&-ia\rho\rho^*\sin\theta/\sqrt{2},\\
\mu&=&\rho^2 \rho^*\triangle/2,\\
\gamma&=&\mu+\rho\rho^*(r-M)/2,\\
\alpha&=&\pi-\beta^*,\\
\psi_2&=&M \rho^3.
\end{eqnarray}
and the differential operators $D=l^\mu\partial_\mu$,
$\Delta=n^\mu\partial_\mu$, $\delta=m^\mu\partial_\mu$, then the
resulting equation is the well known Teukolsky equation,
\begin{eqnarray}
&&
\left[{\left(r^2+a^2\right)^2\over\triangle}-a^2\sin^2\theta\right] 
{\partial^2\psi \over \partial t^2}
+{4Mar\over \triangle} {\partial^2 \psi\over\partial t \partial \varphi} 
+\left[{a^2\over \triangle}-{1 \over\sin^2\theta}\right] 
{\partial^2 \psi \over \partial \varphi^2}
\nonumber\\
&&
-\triangle^2 {\partial \over \partial r} \left( {1 \over \triangle} 
{\partial \psi \over \partial r}\right) -{1 \over \sin\theta}
{\partial \over \partial\theta} \left(\sin\theta {\partial \psi \over 
\partial \theta}\right) 
+4 \left[{M(r^2-a^2)\over \triangle} -r -ia \cos\theta\right] 
{\partial \psi\over \partial t} 
\nonumber\\
&& 
+4 \left[{a(r-M)\over \triangle}+{i \cos\theta\over \sin\theta}\right]
{\partial\psi\over\partial\varphi}
+(4\cot^2\theta+2)\psi=0,
\end{eqnarray}
where $\psi=
{(r-ia\cos\theta)^4}{\psi_4}$. 
As was discussed in references \cite{CaLoBaKhPu}, one can 
write the initial data for the above equation in terms of the perturbative 
three metric and extrinsic curvature. The formulae for the 
Teukolsky function and its time derivative are,
\begin{eqnarray}
\psi _4 &=&-\left[ {R}_{ijkl}+2K_{i[k}K_{l]j}\right] _{(1)}n^i%
{m}^jn^k{m}^l+8\left[ K_{j[k,l]}+{%
\Gamma }_{j[k}^pK_{l]p}\right] _{(1)}n^{[0}{m}^{j]}n^k{m}^l
\label{psi} \\
&&\ -4\left[ {R}_{jl}-K_{jp}K_l^p+KK_{jl}
\right] _{(1)}n^{[0}{m}^{j]}n^{[0}{m}^{l]}, 
\nonumber\\
\partial _t\psi _4 &=&N_{(0)}^\phi \partial _\phi \left( \psi _4\right) -n^i%
{m}^jn^k{m}^l\left[ {\partial }_0R_{ijkl}\right]
_{(1)}  \label{psipunto} \\
&&+8n^{[0}{m}^{j]}n^k{m}^l\left[ {\partial }
_0K_{j[k,l]}+{\partial }_0\Gamma _{j[k}^pK_{l]p}+{
\Gamma }_{j[k}^p{\partial }_0K_{l]p}\right] _{(1)}  \nonumber \\
&&-4n^{[0}{m}^{j]}n^{[0}{m}^{l]}\left[ {
\partial }_0{R}_{jl}-2K_{(l}^p{\partial }
_0K_{j)p}-2N_{(0)}K_{jp}K_q^pK_l^q\right.  \nonumber \\
&&\left. +K_{jl}{\partial }_0K+K{\partial }_0K_{jl}\right] _{(1)} \nonumber \\
&& +2\{\psi_4 (l_i \Delta -m_i\bar{\delta}) N^{i~(0)}
+\psi_3 (n_i\bar{\delta}-\bar{m}_i \Delta) N^{i~(0)}\},\nonumber
\end{eqnarray}
where
\begin{eqnarray}
\psi _3 &=&-\left[ {R}_{ijkl}+2K_{i[k}K_{l]j}\right] _{(1)}
l^i{n}^j\bar{m}^k{n}^l+4\left[ K_{j[k,l]}+{
\Gamma }_{j[k}^pK_{l]p}\right] _{(1)}
(l^{[0}{n}^{j]}\bar{m}^k{n}^l-n^{[0}\bar{m}^{j]}l^k{n}^l)
\label{psi3} \\
&&-2\left[ {R}_{jl}-K_{jp}K_l^p+KK_{jl}
\right] _{(1)}
(l^{[0}{n}^{j]}\bar{m}^0{n}^{l}-l^{[0}{n}^{j]}n^0\bar{m}^{l}), 
\nonumber
\end{eqnarray}
$N_{(0)}=(g_{\text{kerr}}^{tt})^{-1/2}$ is the zeroth order lapse, 
$n^i$ in these equations should be taken to be related to that of the
original tetrad as $n^0= N_{(0)}n^0_{\rm orig}, 
n^i=n^i_{\rm orig}+N^{i~(0)}n^0$.
Latin indices run from 1 to 3, and the brackets are
computed to only first order (zeroth order excluded). The derivatives involved
in the above expressions can be computed in terms of the initial data on 
the Cauchy hypersurface as,
\begin{equation}
{\partial }_0K=N_{(0)}K_{pq}K^{pq}-{\nabla }%
^2N_{(0)},  \label{Kpunto}
\end{equation}
\begin{equation}
{\partial }_0{R}=2K^{pq}{\partial }%
_0K_{pq}+4N_{(0)}K_{pq}K_s^pK^{sq}-2K{\partial }_0K,
\label{Rpunto}
\end{equation}
\begin{eqnarray}
{\partial }_0R_{ijkl} &=&-4N_{(0)}\left\{ K_{i[k}{R}%
_{l]j}-K_{j[k}{R}_{l]i}-\frac 12{R}\left(
K_{i[k}g_{l]j}-K_{j[k}g_{l]i}\right) \right\}  \label{Rijklpunto} \\
&&\ +2g_{i[k}{\partial }_0{R}_{l]j}-2g_{j[k}{%
\partial }_0{R}_{l]i}-g_{i[k}g_{l]j}{\partial }_0%
{R}+2K_{i[k}{\partial }_0K_{l]j}-2K_{j[k}{%
\partial }_0K_{l]i},  \nonumber
\end{eqnarray}
and,
\begin{equation}
{\partial }_0K_{ij}=N_{(0)}\left[ \bar R%
_{ij}+KK_{ij}-2K_{ip}K^p{}_j-N_{(0)}^{-1}\bar \nabla _i\bar \nabla
_jN_{(0)}\right] _{(1)}.  \label{Kijpunto}
\end{equation}

Remarkably, the above formulae are coordinate independent! Therefore
the only adjustment needed to specify initial data for the evolution
equations we will derive in the next two sections is to insert the
appropriate background quantities in the above formulae.

\section{The Teukolsky equation in Kerr--Schild coordinates:\\
Ingoing Eddington Finkelstein coordinates}
\label{ingoing}

The initial data proposed by \cite{HuMaSh,winicour} is constructed using
ingoing Eddington--Finkelstein (IEF) coordinates (strictly speaking, since
the initial data might include net angular momentum, one is really talking
about the generalization of IEF coordinates to the rotating case, 
commonly referred to as Kerr coordinates).  This is in part due to
the fact that these families of initial data are currently being
evolved using a numerical code where the black holes are treated using
the ``excision'' technique. This technique requires coordinates that
penetrate the horizon, such as the IEF ones. The IEF coordinates
$(\tilde{V},r,\theta,\varphi)$ for the 
Kerr metric are defined through a redefinition of the time coordinate 
of the Boyer--Lindquist coordinates as,
\begin{eqnarray}
\tilde{V}=t+r^*\\
\tilde{\varphi}=\varphi+\int {a\over \triangle} dr
\end{eqnarray}
where $r^*$ is the natural generalization to the Kerr case of the
usual Schwarzschild ``tortoise'' coordinate, and is defined by,
\begin{equation}
r^*=\int {r^2+a^2\over r^2-2 M r +a^2}dr. \label{rstar}
\end{equation}

The codes currently being used to evolve the initial data in IEF are
written in terms of a coordinate $\tilde{t}=\tilde{V}-r$. We will
therefore derive the Teukolsky equation in the
$(\tilde{t},r,\theta,\tilde{\varphi})$ coordinates.
The Kerr metric in these coordinates,
\begin{eqnarray}
ds^2=&&\left(1-2Mr/\Sigma\right) d\tilde{t}^2-\left(1+2Mr/\Sigma\right)dr^2 
-\Sigma d\theta^2
-\sin^2\theta\left(r^2+a^2+2Ma^2r\sin^2\theta/\Sigma\right) d\tilde{\varphi}^2
\nonumber \\
&&- \left(4Mr/\Sigma\right)d\tilde{t}dr
+ \left(4Mra\sin^2\theta/\Sigma\right)d\tilde{t}d\tilde{\varphi}
+2a \sin^2\theta\left(1+2 M r/\Sigma\right)d\tilde{r}d\tilde{\varphi},
\label{KSmetric}
\end{eqnarray}
In addition to changing coordinates, it is immediate to see that one
needs also to change tetrads. The usual Kinnersley tetrad is singular
at the horizon, and therefore leads to a Teukolsky equation that is singular.
However, it is easy to fix this problem by just
rescaling $l^\mu$ by a factor of $\triangle$ and dividing $n^\mu$ by 
$\Delta$. This does not change the orthogonality properties of the
tetrad, but makes it well defined \footnote{We wish to thank Steve
Fairhurst and Badri Krishnan for suggesting this rescaling.}.
Therefore we re-derive the Teukolsky equation using as new tetrad
vectors,
\begin{eqnarray}
l^\mu &=&[\triangle+4Mr,\triangle,0,2 a] \\ 
n^\mu &=&[{1\over 2\Sigma},-{1  \over 2\Sigma},0,0]
\end{eqnarray}

This redefinition of the tetrad vectors changes the values of the 
Newman--Penrose scalars. The new values are,
\begin{eqnarray}
\epsilon &=& r-M\\
\gamma &=& \mu = -{1 \over 2} {r + i a \cos\theta\over \Sigma^2}\\
\rho &=& -(r+i a \cos\theta) {\triangle \over \Sigma}
\end{eqnarray}
with $\alpha,\beta,\pi,\tau,\psi_2$ remain unchanged. The resulting
Teukolsky equation reads,
\begin{eqnarray}
&&
{1 \over 2} {\triangle+4 M r \over \Sigma}
{\partial^2 \psi_4 \over \partial \tilde{t}^2} 
-{1 \over 2} {\triangle\over \Sigma}
{\partial^2 \psi_4 \over \partial r^2} 
-2 {Mr \over \Sigma} {\partial^2 \psi_4 \over \partial r \partial \tilde{t}}
- {a \over \Sigma} {\partial^2 \psi_4 \over \partial r \partial \tilde\varphi}
-{1 \over 2} {1 \over \Sigma}
{\partial^2 \psi_4 \over \partial \theta^2}
-{1 \over 2} {1 \over \sin^2\theta\Sigma}
{\partial^2 \psi_4 \over \partial \tilde\varphi^2}
\nonumber\\
&&
-\left\{{3a^2(r-M)\cos^2\theta+r(7r^2+4a^2-11Mr)+4ia\cos\theta\triangle
\over \Sigma^2}\right\}{\partial \psi_4\over \partial r}
\nonumber\\
&&
-\left\{{r^2(2r+11M)+a^2(2r-3M)\cos^2\theta
-2ia(r^2+7Mr+a^2\cos^2\theta)\cos\theta
\over\Sigma(r-i a \cos\theta)^2}\right\}{\partial \psi_4\over \partial t}
\nonumber\\
&&
-\left\{{(r^2+8a^2-9a^2\cos^2\theta)\cos\theta +2iar(4-5\cos^2\theta)\over
2\Sigma (r-i a \cos\theta)^2\sin\theta}\right\}
{\partial \psi_4\over\partial\theta}
\nonumber\\
&&
-\left\{{4ar(\sin^2\theta-\cos^2\theta)-2i
(r^2+2a^2-3a^2\cos^2\theta)\cos\theta
\over \Sigma(r-ia\cos\theta)^{2}\sin^{2}\theta}\right\}
{\partial \psi_4\over\partial\tilde\varphi}
\nonumber\\
&&
+\left\{{24Mr\sin^2\theta-19r^2+(21r^2-9a^2+7a^2\cos^2\theta)\cos^2\theta
-2ia [(7r-6M)\cos^2\theta-(5r-6M)]\cos\theta
\over \Sigma(r-ia\cos\theta)^{2}\sin^{2}\theta}\right\}\psi_{4} =0.
\end{eqnarray}

This equation  can be used to study a variety
of different things, including boundary conditions at the excision
region (inside the horizon), perturbations inside and near the horizon,
and most significantly, we could use this implementation to compare
and perhaps even continue evolutions from full numerical codes that use
Kerr--Schild coordinates.

The Penetrating Teukolsky Code (PTC) evolves the following equation, where $\psi = (r-ia\cos(\theta))^{4} \psi_4$ is the Teukolsky function:
\begin{eqnarray}
&&
(\Sigma + 2Mr){{\partial^2 \psi}\over 
{\partial \tilde t^2}} - \triangle {{\partial^2 \psi}\over 
{\partial r^2}} - 6(r - M){{\partial \psi}\over
{\partial r}}
\\
&&
-{{1}\over {\sin \theta}}{{\partial}\over {\partial \theta}} \left (
\sin \theta {{\partial \psi}\over {\partial \theta}}\right ) -{{1}\over
{\sin^2 \theta}}{{\partial^2 \psi}\over {\partial \tilde \varphi^2}} 
-4Mr{{\partial^2 \psi}\over {\partial \tilde t \partial r}}
-2a {{\partial^2 \psi}\over {\partial r\partial \tilde \varphi}} 
\nonumber\\
&&
+ \left ({4i\cot\theta \over \sin\theta}
\right ) {{\partial \psi}\over {\partial \tilde \varphi}}\nonumber\\
&& 
- \left ({4r+4ia\cos\theta+6M}\right ) {{\partial \psi}\over {\partial
\tilde t}} 
 + 2(3\cot^{2}\theta-\csc^{2}\theta)\psi = 0.
\nonumber
\end{eqnarray}

To implement this equation numerically, we break the above equation down
into a 2+1 dimensional one, using a decomposition of the Teukolsky
function into angular modes, $\psi=\Sigma \psi_{m} e^{im {\tilde \varphi}}$. 
The Teukolsky
equation for each $m$ mode now looks like the following:
\begin{eqnarray}
&&
(\Sigma + 2Mr){{\partial^2 {\psi_m}}\over 
{\partial \tilde t^2}} - \triangle {{\partial^2 {\psi_m}}\over 
{\partial r^2}} - (2aim+6r - 6M){{\partial {\psi_m}}\over
{\partial r}}
\\
&&
-{{1}\over {\sin \theta}}{{\partial}\over {\partial \theta}} \left (
\sin \theta {{\partial {\psi_m}}\over {\partial \theta}}\right ) 
-4Mr{{\partial^2 {\psi_m}}\over {\partial \tilde t \partial r}}
- \left ({4r+4ia\cos\theta+6M}\right ) {{\partial {\psi_m}}\over {\partial
\tilde t}} 
\nonumber\\
&&
 + (4\cot^{2}\theta-2+{m^2}\csc^{2}\theta-4m\cot\theta\csc\theta){\psi_m} 
= 0.
\nonumber
\end{eqnarray}
We use Lax-Wendroff technique to numerically implement these simplified
set of equations exactly as done in \cite{AnKrLaPa}. A full discussion of
the applications of this code will be presented elsewhere. Here we just
highlight in the figures how the code indeed evolves perturbations inside
and outside the horizon as expected.

\begin{figure}[h]
\centerline{\psfig{figure=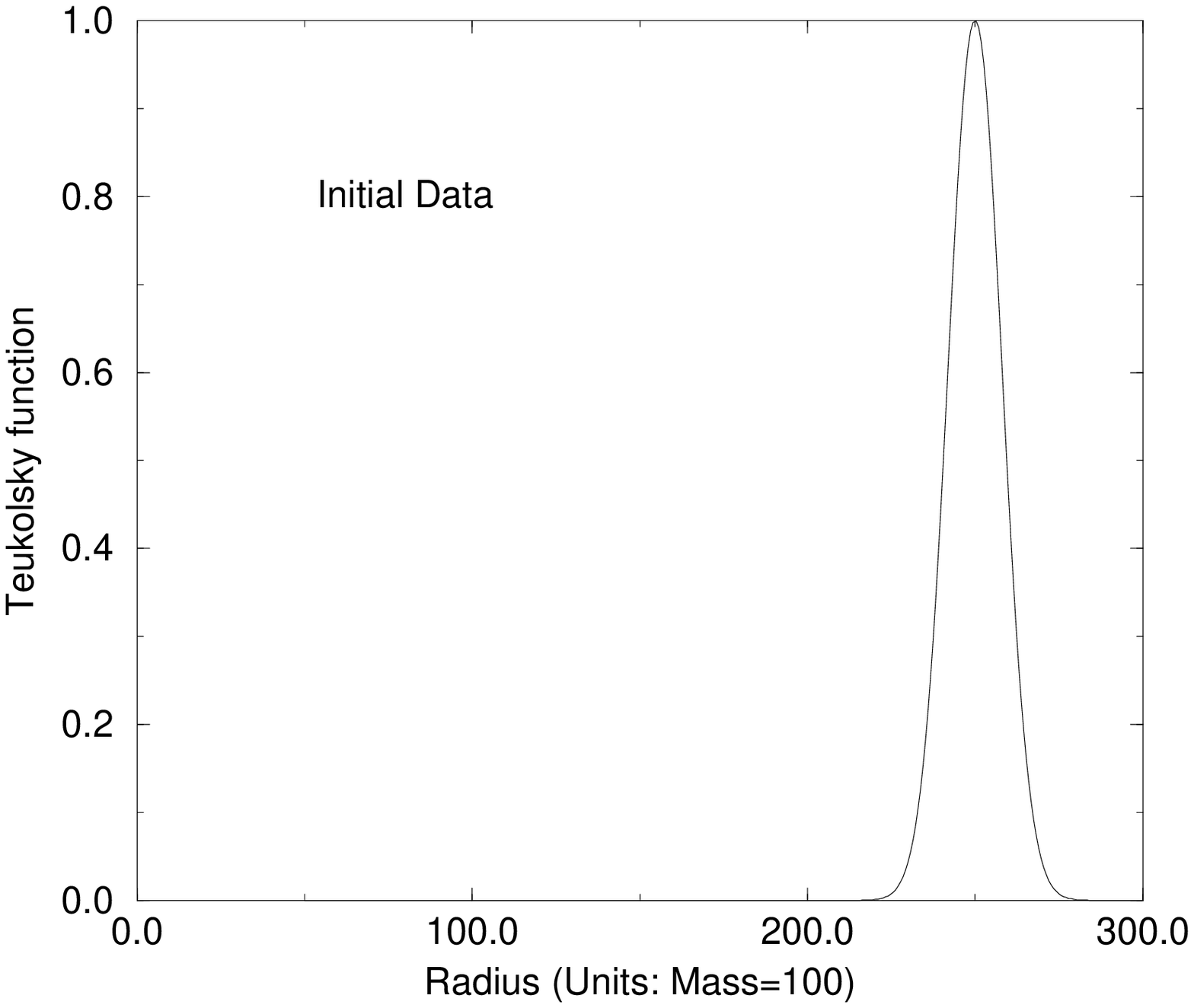,width=70mm,height=50mm}}
\centerline{\psfig{figure=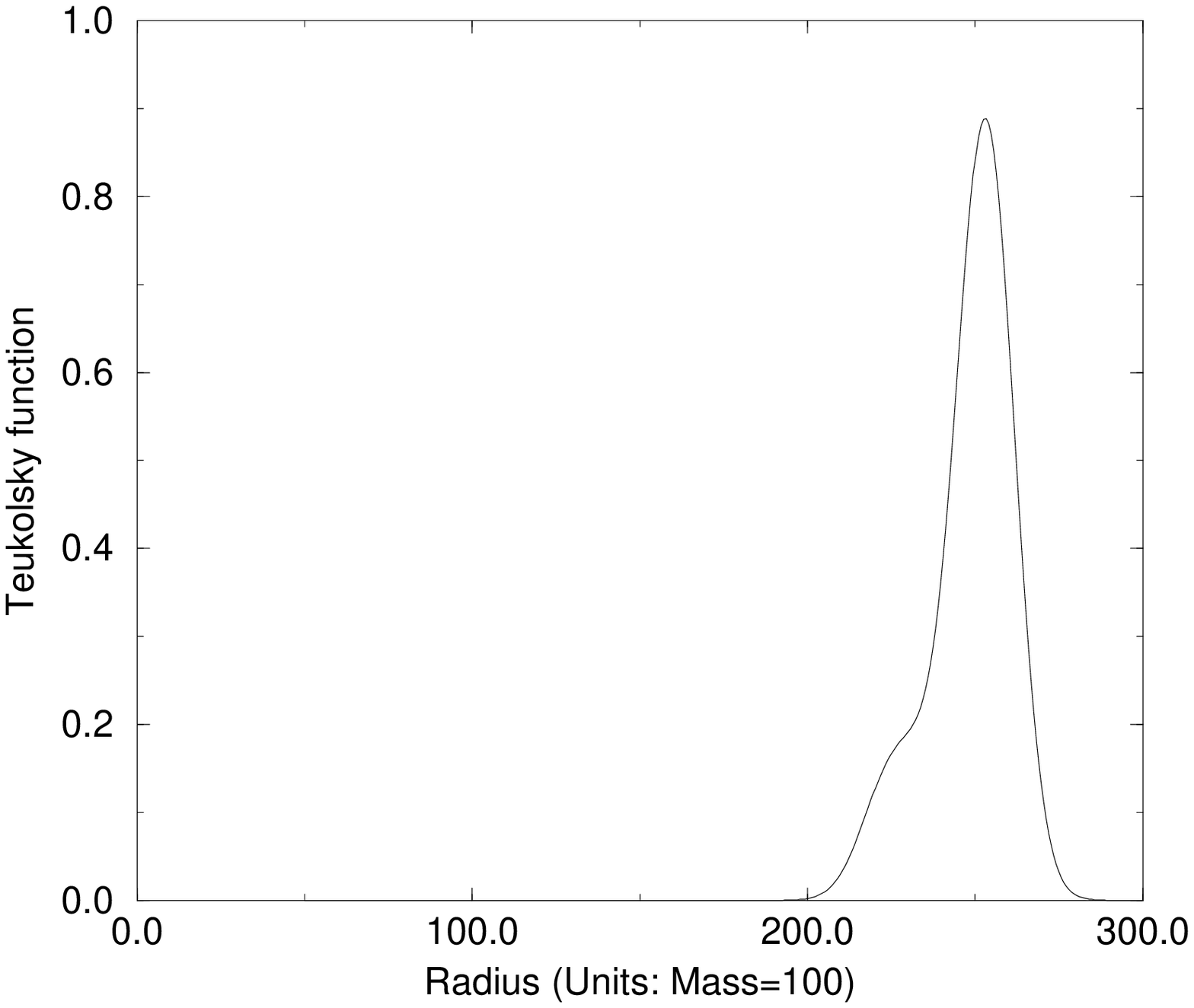,width=70mm,height=50mm}}
\centerline{\psfig{figure=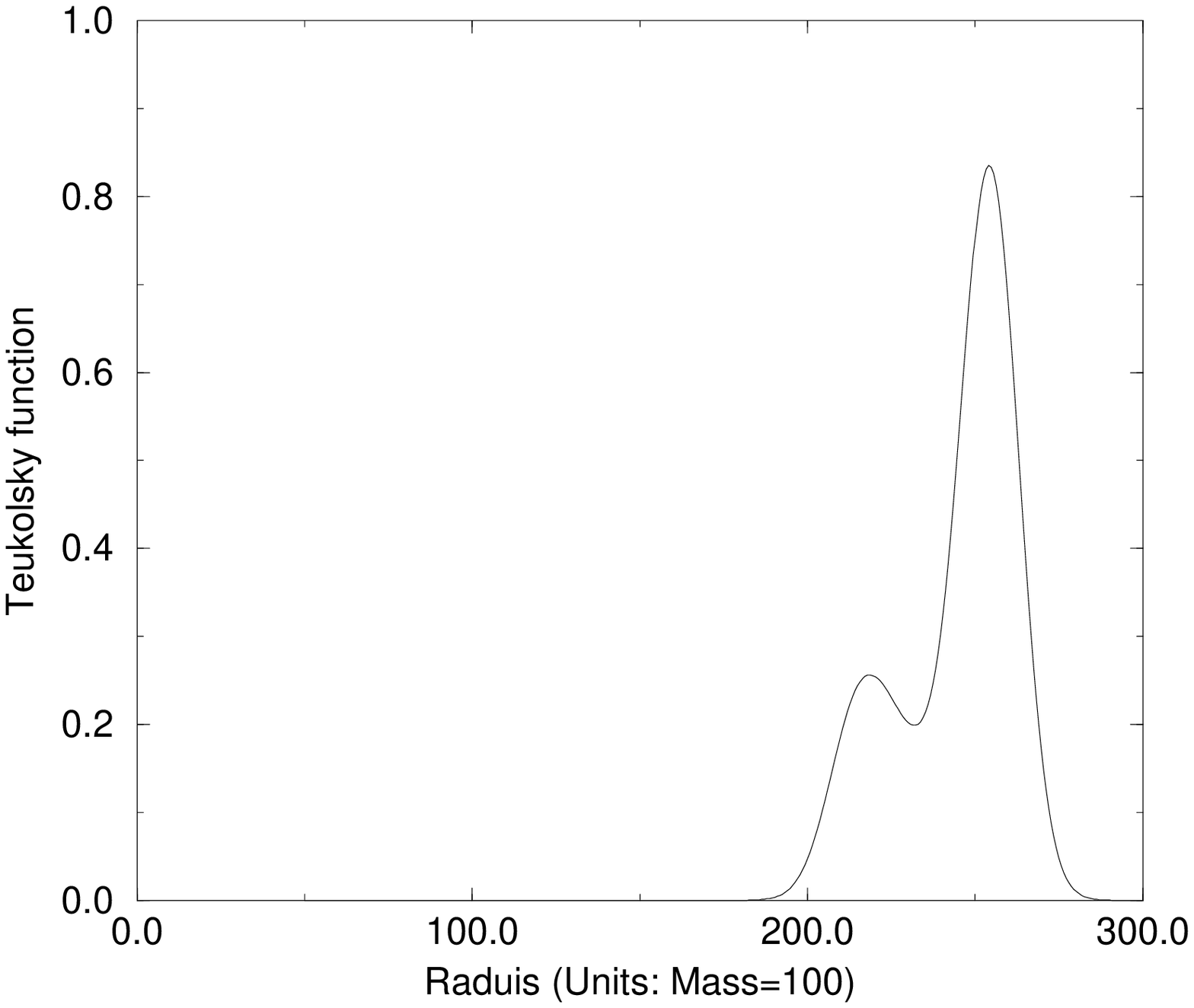,width=70mm,height=50mm}}
\centerline{\psfig{figure=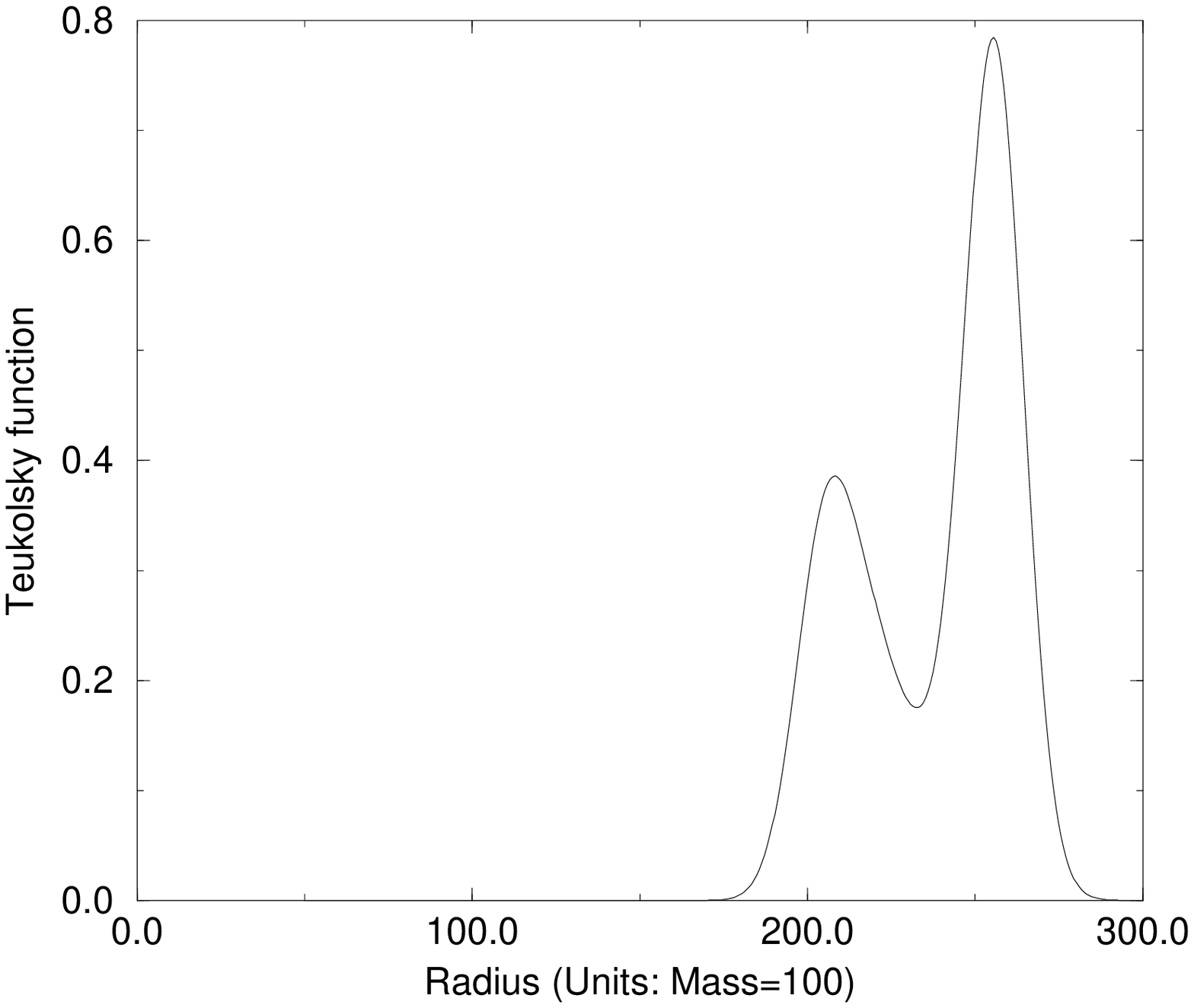,width=70mm,height=50mm}}
\caption{Numerical evolution of the horizon penetrating Teukolsky
equation. The horizon is at $r=200$. We give as initial data a Gaussian pulse
outside the horizon. The background black hole has a mass of $100$ units, and
zero spin.
Shown above are snapshots of the evolution of the $m=0$ mode 
at $t=30$, $40$, and $50$
units. The pulse splits into two pulses, one infalling
towards the singularity at $r=0$, moving smoothly through the
horizon. The other pulse moves rightwards and eventually escapes to
scri. Notice that the speed of light does not appear as a $t=r$ 
motion in these coordinates!}
\end{figure}

\section{Conclusions}

We have set up the black hole perturbation framework in Kerr--Schild
type coordinates. This framework will be useful for comparisons with
fully nonlinear numerical codes currently being implemented that run
naturally in Kerr--Schild coordinates. We have
also discussed the numerical implementation of the perturbative
evolution equation and how to set up initial data in terms of the
initial value data that will be available for black hole
collisions. Implementation of this framework for numerical
computations is essentially complete.

This formalism allows us to study in a natural way perturbations
close to the horizon and may also be of interest to study and test in a
concrete fashion several attractive properties of ``isolated
horizons'' \cite{isho}. This formalism allows us to make several predictions
about quantities defined with notions intrinsic to the black hole
(like a concept of local mass and angular momentum), and their evolution.
Several attractive formulae, for instance relating the ADM mass to the
``local horizon mass'' and the ``radiation content'' can be worked out.
Having a perturbative formalism that operates correctly near and on the
horizon will allow us to test the validity of these formulae. This issue
is currently under study. 

\section{Acknowledgments}
We wish to thank Dar\'{\i}o N\'u\~nez for pointing out several typos in
an earlier version of this paper.
This work was supported in part by grants, 
NSF-INT-9722514, 
NSF-PHY-9423950,
NSF-PHY-9800973, 
NSF-PHY-9800970,
NSF-PHY-9800973,
and by funds of the Pennsylvania State University. 
M.C. holds a Marie Curie Fellowship (HPMF-CT-1999-00334).

\newpage

\appendix{\bf Appendix:The Teukolsky equation in Outgoing Eddington--Finkelstein coordinates:}
\\

The outgoing Eddington--Finkelstein coordinate form of the Kerr metric
might be useful to evolve in a perturbative fashion the outgoing wave
zone exterior, say, to a tube in which a Cauchy evolution code is
used, as in the spirit of \cite{grandlychallenged2}. The outgoing
Eddington--Finkelstein coordinates for the Kerr metric are obtained
by introducing a time variable,
\begin{eqnarray}
\tilde{U}&=&t-r^*\\
\tilde{\varphi}&=&\varphi-\int {a\over \triangle} dr
\end{eqnarray}
where $r^*$ is defined as in (\ref{rstar}). 

As in the previous section, one presumably wishes to write a Cauchy 
evolution code for the resulting perturbative equation. It is therefore
appropriate to introduce a time coordinate,
\begin{equation}
\tilde{t}= \tilde{U}+r,
\end{equation}
and consider the Teukolsky equation in $(\tilde{t}, r,
\theta,\tilde{\varphi})$ coordinates.
The Kerr metric in these coordinates is the same as 
\ (\protect\ref{KSmetric}) except an opposite sign in the 
$drd\tilde{t}$ term.

The Newman--Penrose scalars do not change. The differential operators do.
The Kinnersley tetrad in the new coordinates is,
\begin{eqnarray}
l^\mu &=&[1,1,0,0]\\
n^\mu &=&[{\triangle \over 2\Sigma}\left(1+{4M r \over \triangle}\right),
-{\triangle \over 2\Sigma},0,{a \over \Sigma}]\\
m^\mu &=&[i a \sin \theta,0,1,{i \over \sin\theta}]/
(\sqrt{2}(r+ia\cos\theta)),
\end{eqnarray}

The resulting Teukolsky equation in these coordinates is given by,
\begin{eqnarray}
&&
\left(\Sigma+2Mr\right){\partial^2 \psi \over \partial \tilde t^2}
-\triangle {\partial^2 \psi \over \partial r^2}
-{\partial^2 \psi \over \partial \theta^2}
-{1 \over \sin^2\theta} {\partial^2 \psi \over \partial \tilde \varphi^2}
+4Mr {\partial^2 \psi \over \partial \tilde t\partial r}
+2 a {\partial^2 \psi \over \partial r \partial \tilde \varphi}
\nonumber\\
&&
-2\left(2r +M+2 i a \cos\theta\right) {\partial \psi \over \partial \tilde t}
+2\left(r-M\right) {\partial \psi \over \partial r}
-\cot\theta {\partial \psi \over \partial \theta}
+4i {\cos\theta\over\sin^2\theta}
{\partial \psi \over \partial \tilde\varphi}
+\left(2+4 \cot^2\theta\right) \psi =0,
\end{eqnarray}

which can be rewritten in a form more reminiscent of the ordinary 
Teukolsky equation as,
\begin{eqnarray}
&&
(\Sigma + 2Mr){{\partial^2 \psi}\over{\partial \tilde t^2}} 
- {{\partial}\over {\partial r}}\left (\triangle
{{\partial \psi}\over {\partial r}} \right ) 
+ 4(r - M){{\partial \psi}\over {\partial r}}\\
&&
-{{1}\over {\sin \theta}}{{\partial}\over {\partial \theta}} \left (
\sin \theta {{\partial \psi}\over {\partial \theta}}\right ) 
-{{1}\over{\sin^2 \theta}}{{\partial^2 \psi}\over {\partial \tilde \varphi^2}}
-2a {{\partial^2 \psi}\over {\partial \tilde \varphi \partial r}} 
-4Mr{{\partial^2 \psi}\over {\partial \tilde t \partial r}} 
\nonumber\\
&&
-2\left(2r +M+2 i a \cos\theta\right) 
{{\partial \psi}\over {\partial \tilde t}}
+4i {\cos\theta\over\sin^2\theta}
{{\partial \psi}\over {\partial \tilde \varphi}} 
 + 2(1 + 2\cot^2 \theta)\psi = 0,
\nonumber
\end{eqnarray}
where again $\psi={(r-ia\cos\theta)^4}{\psi_4}$, 
and we see that the main differences with the equation we derived in
the previous section are: a) the sign of the terms with the mixed
$r,\tilde\varphi$ and $r,\tilde t$ derivatives, b) the first order
derivative in $\tilde t$ and $\tilde\varphi$ terms. 

As in the previous cases, the generic initial data formulae we presented
are still valid in this case.

\end{document}